**Ballistic Phonon Thermal Transport in Multi-Walled Carbon Nanotubes**

H.-Y. Chiu[†], V. V. Deshpande[†], H. W. Ch. Postma, C. N. Lau[1], C. Mikó[2], L. Forró[2], M. Bockrath[*]

We report electrical transport experiments using the phenomenon of electrical breakdown to perform thermometry that probe the thermal properties of individual multi-walled nanotubes. Our results show that nanotubes can readily conduct heat by ballistic phonon propagation. We determine the thermal conductance quantum, the ultimate limit to thermal conductance for a single phonon channel, and find good agreement with theoretical calculations. Moreover, our results suggest a breakdown mechanism of thermally activated C-C bond breaking coupled with the electrical stress of carrying $\sim 10^{12}$ A/m$^2$. We also demonstrate a current-driven self-heating technique to improve the conductance of nanotube devices dramatically.

Department of Applied Physics, California Institute of Technology, Pasadena, CA 91125

[†]These authors contributed equally to this work

[1]Department of Physics, University of California, Riverside CA 92521

[2]IPMC/SB, EPFL, CH-1015 Lausanne-EPFL, Switzerland

[*] To whom correspondence should be addressed. E-mail: mwb@caltech.edu

The ultimate thermal conductance attainable by any conductor below its Debye temperature is determined by the thermal conductance quantum[1, 2]. In practice, phonon scattering reduces the thermal conductivity, making it difficult to observe quantum thermal phenomena except at ultra-low temperatures[3]. Carbon nanotubes have remarkable thermal properties[4-7], including conductivity as high as ~3000 W/m-K[8]. However, phonon scattering still has limited the conductivity in nanotubes.

Here we report the first observation of ballistic phonon propagation in micron-scale nanotube devices, reaching the universal limit to thermal transport. In this qualitatively different regime, quantum mechanics limits the entropy flow, giving a maximum thermal conductance and an absolute physical limit to the information bandwidth that a nanotube can transport per unit power[1, 2].

Our experimental approach is to heat multi-walled nanotubes (MWNTs) with an electrical current and monitor temperature by the electrical breakdown phenomenon[9]. From careful analysis of our data, we obtain an experimental measurement of the thermal conductance quantum, which agrees with theoretical predictions[1, 2, 10] as well as thermal transport results on cryogenically-cooled $Si_3N_4$ nanobridges[3]. This demonstrates that fundamental knowledge about thermal transport in nanotubes can be obtained from an electrical transport experiment. This knowledge, which is challenging to obtain by other means, also contributes toward understanding thermal management issues relevant to the rational design of nanotube interconnects and logic devices.

Furthermore, our data suggest a shell breakdown mechanism based on thermal activation of σ-π* electronic transitions, similar to the electronic excitation of such transitions in scanning tunneling microscope (STM) cutting experiments[11, 12]. Finally, we demonstrate an annealing process whereby nanotube device conductance is improved dramatically by an electrical current, which is likely related to the nanotube self-heating. This greatly improves the yield of high-conductance nanotube devices.

Fabrication of free-standing and substrate-supported MWNT devices was described elsewhere(*13, 14*). A schematic device diagram is shown in Fig. 1A and a scanning electron microscope (SEM) device image in Fig. 1B. We mainly studied devices in which electrical leads were placed over the tubes. In the work of Collins *et al.*(*9*), it was shown that sufficiently high electrical power dissipation in MWNTs causes the current *I* to drop in abrupt events separated by ~1 s, due to the ablation of individual nanotube shells. It was carefully argued that the breakdown temperature $T_B$ was ~900 K(*9*). Nevertheless, Joule heating alone is not likely to account entirely for the shell breakdown(*15*).

To address the role of $T_B$ in the breakdown process, we compared the behavior of both freestanding devices and supported nanotube devices. In the latter, the substrate provides an additional cooling pathway for the nanotubes. Figure 1C shows current-voltage (*IV*) data from three freestanding nanotube devices with radius *R*=10 nm, determined by SEM imaging(*16*). The samples' lengths *L* were 0.50 μm, 0.64 μm, and 1.58 μm, respectively (top to bottom). The *IV* curves end at an abrupt ~10 μA current drop, marked by the black circles, indicating shell breakdown. The voltage *V* was then quickly reduced, to prevent further shell breakdown(*17*). The dotted line is a constant power curve in the *IV* plane corresponding to the breakdown power *P* for the shortest tube. From this, we deduce that increasing *L* leads to decreasing *P*.

Figure 1D shows *IV*s from supported nanotube data with *L*= 0.74 μm, 1.26 μm, and 1.66 μm, for curves with ends going from left to right, respectively. *P* increases with *L*, showing the opposite behavior from free-standing nanotubes, and is relatively insensitive to $R$, which ranged from 8-14 nm.

Figure 1E shows a log-log plot of $P$ *vs.* $L$ for five free-standing tubes (open circles) and five supported tubes (filled squares). The open circles approximately follow the dashed line, exhibiting $P \propto L^{-1}$ behavior, while the filled squares follow the dotted line, exhibiting $P \propto L$ behavior.

This behavior can be understood using Fick's law, where the thermal current in the nanotube and substrate is proportional to the temperature gradient. For freestanding tubes, the necessary power $P$ to increase the temperature at the tube center by $\Delta T$ is then

$$P = 8\pi R^2 \kappa \Delta T / L, \tag{1}$$

where $\kappa$ is the characteristic nanotube thermal conductivity(*18*). Taking the previous estimate $T_B \sim 900$ K(*9*), a linear fit to our data with $\Delta T = 600$ K (dashed line in Fig. 1E) yields a thermal conductivity of $\kappa \approx 600$ W/m·K, consistent with previous thermal conductivity measurements on individual MWNTs in the diffusive regime(*8*). For supported nanotubes, the relation $P \propto L$ indicates that the cooling occurs mainly by heat conduction into the substrate. We estimate heat transport in this geometry as between concentric cylinders. This yields $P = 2\pi L \kappa_s T / \ln(R_0/R)$, with $R_0$ the outer cylinder radius at which $T$ drops to the ambient value, and $\kappa_s$ the substrate thermal conductivity. Taking $R_0 = 50$ nm, and $R = 10$ nm the fit shown by the dotted line in Fig. 1E yields $\kappa_s \sim 0.5$ W/m·K, in agreement with the bulk thermal conductivity of $SiO_2$, $\kappa_s \approx 1.5$ W/m·K. The opposite scaling behavior of $P$ *vs.* $L$ between the supported and free-standing nanotube devices, taken together with the quantitative agreement obtained with the expected values for $\kappa$ and $\kappa_s$ demonstrate that the shell ablation occurs at a well-defined temperature $T_B$.

We now focus exclusively on freestanding nanotube devices, representing a broad range of $L$ and $R$ values. Figure 2 shows $P/8\pi R^2$ *vs.* $L^{-1}$ for ~30 samples. Based eq. 1, we expect plotting the normalized power $P_N = P/8\pi R^2$ *vs.* $L^{-1}$ should yield a straight line with a slope of $T_B\kappa$. Remarkably, although the initial trend for the longer tubes (open circles) follows the straight dashed line, for nanotubes with $L^{-1} \gtrsim (0.5\mu m)^{-1}$ (filled squares) $P_N$ saturates and becomes $L$-independent. This indicates that heat flow exiting the nanotube occurs at an $L$-independent rate, depending *only* on $R$. The remainder of the shells are then broken, producing a gap in the nanotube. The Fig. 2 inset shows the gap position, normalized to the suspended tube length. This breakpoint is near the center for tubes longer than ~0.5 μm, but for $L^{-1}$ in the saturation regime the scatter in the breakpoint values increases. This suggests the temperature distribution along the tube becomes more uniform.

Figure 3A shows $P$ *vs.* $R$ on a log-log scale. Data from short tube samples (filled squares) fall near a straight line fit to the data in the log-log plot, showing that $P \propto R^\alpha$, with $\alpha = 2.1$. Some data from longer samples (open circles), fall near the line, but for the longest nanotubes, the data points fall below the line. The curve followed by the short nanotube data represents an upper limit to $P$; modulo experimental scatter, data for each nanotube falls on or below the curve and achieves the maximum only for nanotubes with $L \leq 0.5$ μm. The Fig. 3A inset shows the same data and fit on a linear scale.

We now consider potential interpretations for this behavior. One possibility is that a dominant metal-nanotube thermal contact resistance $\kappa_C$ produces the saturation seen in Fig. 2. This is unlikely, however, as $\kappa_C$ was determined to be negligible in ref. (*8*), which also used metal contacts as thermal reservoirs. Furthermore, based on the supported tubes' behavior, we would expect $\kappa_C$, and hence $P$, to be approximately independent of $R$. The observed systematic relationship $P \propto R^2$ differs sharply from this expectation.

Another possibility is that because of ballistic electron transport(*19*) the electrical current primarily heats the electrodes. In this case, the required power to reach $T_B$ may be relatively *L*-insensitive. However, both experiments and theory(*20, 21*) indicate the electronic mean free path due to phonon emission at the high biases applied to our samples is ~10 nm. Since each optical or zone boundary phonon emission is associated with an energy ~180 meV(*21*), we expect even for our shortest nanotube studied (~150 nm) most of the energy $eV$ provided by the electric field to each electron is converted into phonons within the nanotube.

We now discuss the possibility of ballistic *phonon* transport within the nanotube. In this picture, a diffusive heat transport regime with Umklapp inter-phonon scattering as the dominant scattering mechanism(*8*) makes a transition to a ballistic center-of-mass motion regime for sample lengths $L \lesssim 0.5$ μm. This suggests the temperature distribution along the tube should broaden as $L$ decreases, consistent with the data in the Fig. 2 inset. Furthermore, because the characteristic distance the phonons travel before escaping the tube is ~$L/2$, we would infer a characteristic Umklapp scattering mean-free path $l_U$~0.2 μm.

In the ballistic regime, the heat flux carried by the phonons $P_{ph}$ is given by(*3, 10*)

$$P_{ph} = \sum_n \int_{\omega_{n,\min}}^{\omega_{n,\max}} \frac{d\omega_n}{2\pi} \hbar \omega_n [\eta_{ne}(\omega_n) - \eta(\omega_n, T_0)] \xi(\omega_n), \qquad (2)$$

where $\omega_{n,max}$, $\omega_{n,min}$ are the $n$th phonon branch cutoff frequencies, $\eta_{ne}(\omega_n)$ is the non-equilibrium phonon distribution for the $n$th phonon branch, $\eta(\omega_n, T_0)$ is the Bose-Einstein distribution corresponding to the phonons injected into the nanotube at the electrode temperature $T_0$, and $\xi(\omega_n)$ is the transmission coefficient for phonons escaping into the electrodes.

We now make several assumptions to simplify eq. 2. Considering the negligibility of $\kappa_C$ as discussed above, we take $\xi(\omega_n) \approx 1$. Moreover, since $T_B >> T_0$ we neglect $\eta(\omega_n, T_0)$ relative to $\eta_{ne}(\omega_n)$. Since breakdown depends on $P$, rather than $V$ or $I$ separately the hot phonons emitted by the electrons likely achieve thermodynamic equilibrium over a thermalization length $l_{th} << L$ after a few $L$-independent characteristic number of collisions(22). Thus we set $\eta_{ne}(\omega_n) = \eta(\omega_n, T)$, where $T$ is the tube temperature. We also set $\omega_{n,min} \approx nc/R(23)$, where $c \sim 1.5 \times 10^4$ m/s is the in-plane speed of sound in graphite(24). Since $T_B$ is considerably less than $\Theta_D \sim 2500$ K, the graphene Debye temperature, we replace $\omega_{n,max}$ by infinity. Finally, motivated by the observed $P \propto R^2$ relationship we assume that the thermal current is carried by the different nanotube shells in parallel.

Summing over the contribution from each shell in the MWNT spaced by $a$=0.34 nm independently (justified by weak coupling between graphite sheets), the power dissipated by phonons exiting the nanotube is then

$$P_{ph} \approx 2\zeta(3)(k_B T)^3 \frac{R^2}{\pi \hbar^2 ac},$$
(3)

taking into account the heat flow into both contacts and phonon mode degeneracy factor of two, with $\zeta$ the Riemann zeta function. Note that this expression contains no free parameters. We rewrite eq. 3 as $P_{ph}=2M\kappa_Q T$, analogous to the well-known Landauer formula for the ballistic conduction of electrons. Here $M\approx1.5\pi k_B TR^2/hac$, corresponding to the characteristic number of occupied phonon branches, and $\kappa_Q=\pi^2 k_B^2 T/3h$ is the thermal conductance quantum[1, 2].

Plotting $P$ *vs.* $M$ for samples shorter than ~0.5 μm should thus yield a straight line with a slope of $T_B\kappa_Q$. Figure 3B shows such a plot with $T=T_B$=900 K for tubes with $L\lesssim0.5$ μm. The data closely follows a straight-line with a fitted slope of 1.0 μW/branch. From this, we infer a value for the thermal conductance quantum of $6\times10^{-10}$ W/K, in good agreement to the theoretical value given by $\kappa_Q=9\times10^{-10}$ W/K. This is the key experimental finding of this work. Although the accuracy of this measured value of $\kappa_Q$ is somewhat limited by the uncertainty in $T_B$, our experimental determination of $\kappa_Q$ is nevertheless in good quantitative agreement to theory. This demonstrates that we readily reach quantum mechanical limits to thermal transport in our nanotube devices, and that nanotubes can act as ballistic waveguides for phonon propagation. This is the first such observation for a nanostructure above cryogenic temperatures.

Furthermore, our data yields insight into the breakdown mechanism. The well-defined breakdown temperature suggests it requires an initial defect-forming step with activation energy $\Delta$. We expect that the defect formation rate is $\Gamma\sim N\omega_A\exp(-\Delta/k_B T_B)$, where $\omega_A$ is a characteristic attempt frequency, and $N$ is the number of atoms in the tube at temperature $T_B$~900 K. We find $\Delta$~3-4 eV with $\Gamma$=1 s$^{-1}$, a range of $\omega_A$~$10^8$-$10^{18}$ s$^{-1}$ and $N$~$10^6$, highly insensitive to the choice of $N$ and $\Gamma$. One possible origin for $\Delta$ is the formation of a Stone-Wales defect. However, our estimated $\Delta$ is considerably smaller than the theoretically determined barrier ~10 eV to form a Stone-Wales defect in graphite and MWNTs(*25*) ruling out this possible mechanism.

Another, more likely, scenario follows from STM experiments where nanotubes were cut with voltage pulses, showing a well-defined cutting threshold voltage ~3.8 V(*11*). This was interpreted as the signature of an electronic excitation from a $\sigma$ to $\pi^*$ state in the nanotube by the tunneling electrons(*12*). Such transitions produce a local weakening of the carbon-carbon bonds in graphite, with a characteristic energy cost of $E_{\sigma\text{-}\pi^*}$~3.6 eV(see *e.g.*, ref. (*26*)), close to our estimated $\Delta$~3-4 eV. This suggests that in our experiments the available thermal energy provides the energy $E_{\sigma\text{-}\pi^*}$. Combined with the electrical stress of carrying a current density ~$10^{12}$ A/m$^2$, which would readily break metal wires by electromigration(*27*), these $\sigma\text{-}\pi^*$ transitions cause defect formation and a dissipation cascade that ablates the nanotube shell.

Finally, we are able to improve nanotube device conductance considerably using the electric current flow. Freestanding samples with initial low-bias resistance ~50 k$\Omega$ up to 10 M$\Omega$ typically show a rapid increase in conductance as the voltage across the sample is ramped, such as shown in Fig. 4. At higher voltages, a cascade of shell ablation begins and the current decreases in a stepwise fashion. The inset shows that the rise in conductance occurs in a smooth fashion. The increase typically stops when the tube has reached a characteristic temperature ~750 K, as estimated by the applied power at that point, suggesting the phenomenon likely involves nanotube self-heating in conjunction with the current flow. Further experiments, however, are necessary to fully clarify the origins of this behavior, which is of practical value in addressing the challenge of obtaining a high yield of conductive nanotube devices.

In conclusion, our model of ballistic phonon heat transport successfully explains all our experimental observations, including the $L$-independence of $P$ for tubes with $L < 0.5\mu m$, the observed relation $P \propto R^2$, the behavior of the breaking position with $L$, and the quantitative values of $P$. We thereby demonstrate quantum-limited thermal transport in nanotubes, obtaining a thermal conductance quantum value in close agreement with theory. Our results underscore the fantastic capabilities of nanotubes to function as nanoscale thermal conduits under ambient or elevated temperature conditions. They also suggest that optimizing nanotubes' current-carrying capacity will require proper thermal management with an appropriate sample geometry or substrate.

**Fig. 1. (A)** Schematic device diagram showing freestanding multiwalled nanotube on top of electrodes. A voltage $V$ is applied and the current $I$ measured. (**B**) Scanning electron microscope image of individual freestanding nanotube with attached electrodes that were evaporated on top of the tube prior to releasing it using an HF $SiO_2$ etch. (**C**) $IV$ characteristic of three different freestanding nanotube devices with $R$=10 nm. The arrow indicates increasing lengths following. As the suspended nanotube length increases, the power $P$ required to initiate the shell breakdown decreases. Dotted line: iso-power curve at $P$ required to initiate breakdown in the shortest tube. (**D**) $IV$ characteristic of three substrate-supported devices with $R$= 8 nm, 9 nm, and 14 nm that increase in length following the arrow. As the nanotube length increases, the power required to initiate the shell breakdown increases. Dotted line: iso-power curve equal to $P$ required to initiate breakdown in the shortest tube. (**E**) Breakdown power $P$ *vs.* length $L$ on a log-log scale for freestanding tubes (open circles) and substrate supported nanotubes (filled squares). Dotted line: $P \propto L$, dashed line: $P \propto L^{-1}$.

**Fig. 2.** Plot of inverse suspended nanotube length $L^{-1}$ *vs.* normalized power $P/(8\pi R^2)$ for ~30 different nanotube devices. Longer tubes are indicated by open circles, while shorter tubes are indicated by filled squares. Dashed line: linear fit to short tube data, constrained to go through the origin. Inset: breakpoint position normalized to the suspended tube length *vs.* $L^{-1}$.

**Fig. 3. (A)** Log-log plot of $P$ vs. $R$ for ~30 nanotubes devices. Shorter nanotubes are indicated by open circles, while longer nanotubes are indicated by filled squares. Dashed line: linear fit to short nanotube data with a slope $\alpha=2.1$, demonstrating an essentially quadratic relation between $P$ and $R$. Inset: same data and fit, but on a linear scale for comparison. (**B**) $P$ vs. $M$, the characteristic number of occupied phonon branches at temperature $T_B$. Dashed line: linear fit to data taken from nanotubes with $L<0.6$ $\mu$m, constrained to go through the origin.

**Fig. 4.** Current-voltage characteristic of a freestanding nanotube device with $R=14$ nm that has an initial resistance of 50 k$\Omega$. At the voltage indicated by the arrow the current begins to increase dramatically. Inset: expanded view of data in the rapidly increasing region, demonstrating that the current increase occurs in a smooth fashion.

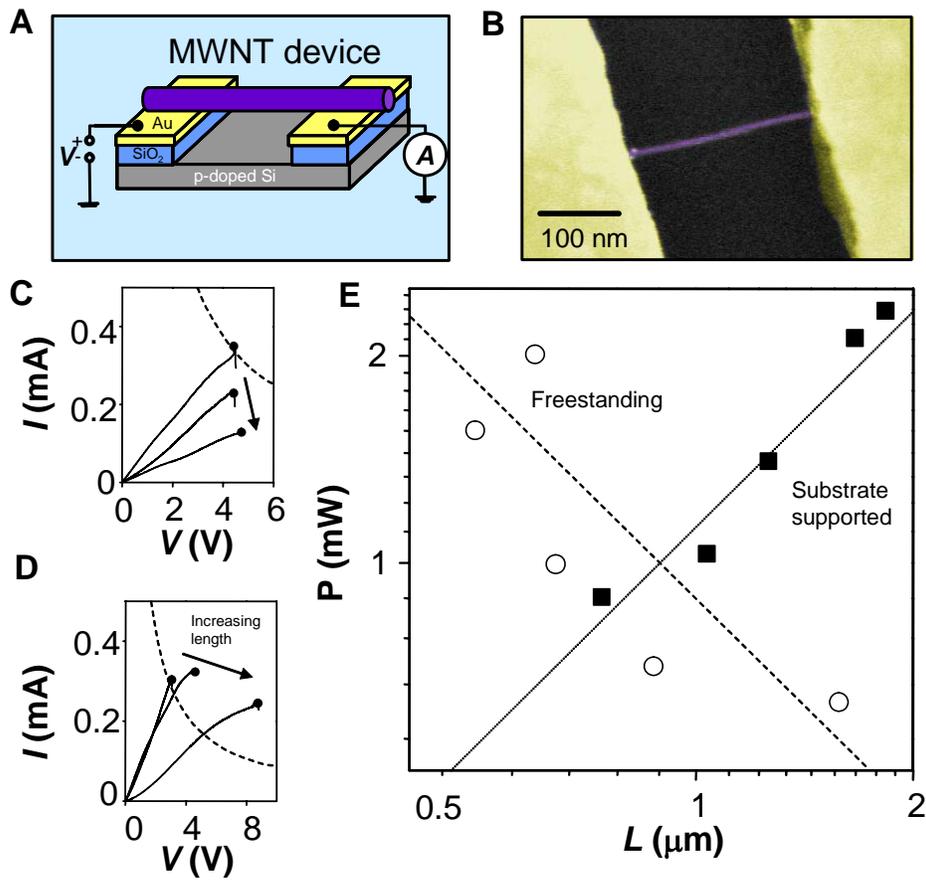

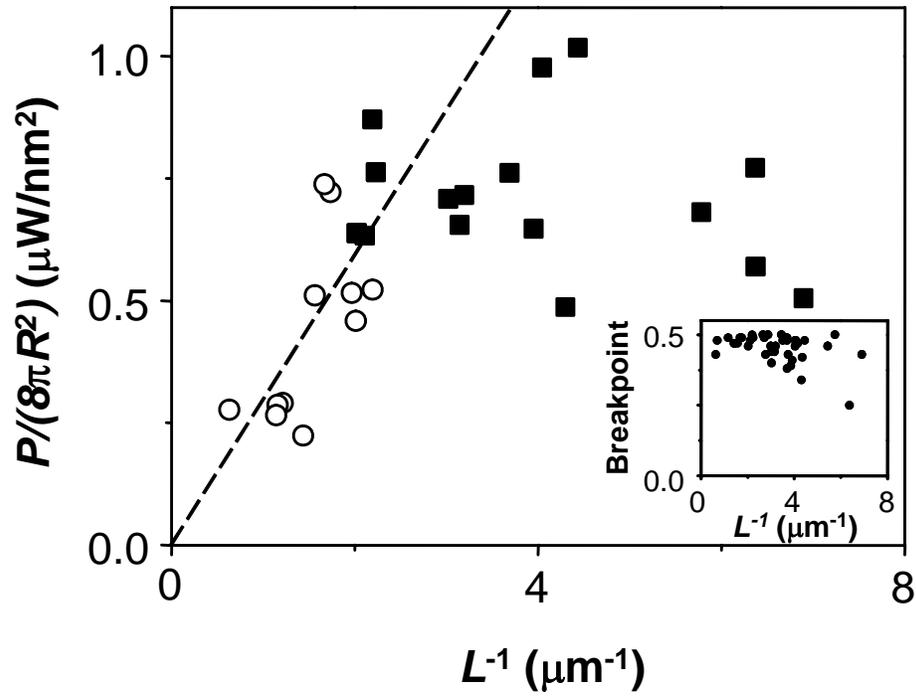

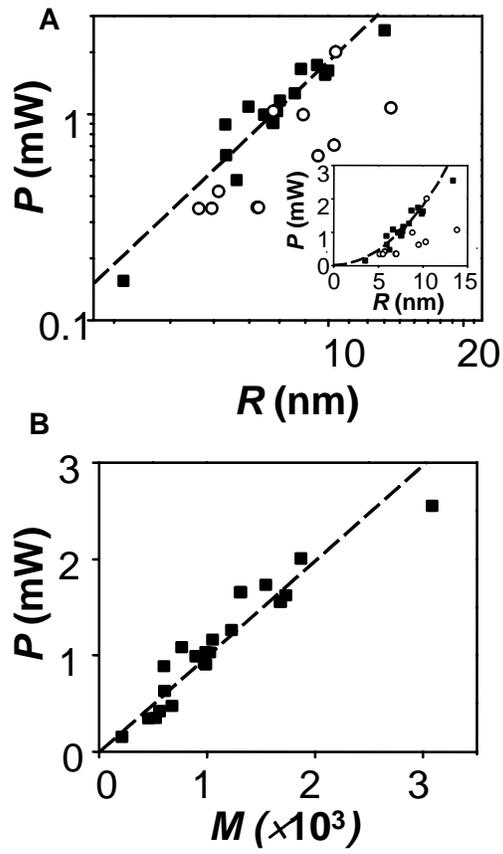

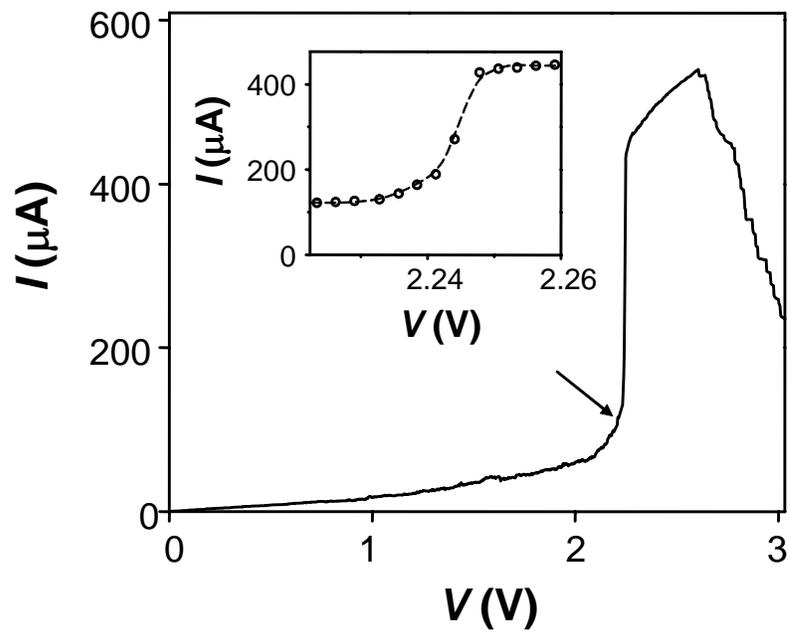